# Room temperature Mott transistor based on resistive switching in disordered $V_2O_3$ films grown on Si

Binoy Krishna De[1,2], V. G. Sathe[1*], Divya[1], Pragati Sharma[1], Shubham Kumar Parate[2], Hemant Singh Kunwar[1], Pavan Nukala[2], and S. B. Roy[1,3]

[1]UGC-DAE Consortium for Scientific Research,
D.A. University Campus, Khandwa Road, Indore-452001, INDIA

[2]Centre for Nano Science and Engineering, Indian Institute of Science, Bangalore-560012, INDIA

[3] Department of Physics, Ramakrishna Mission Vivekananda Education and Research Institute, Belur Math, Howrah, 711202, West Bengal, India.

**Electric field induced giant resistive switching triggered by insulator to metal transition (IMT) is one of the promising approaches for developing a new class of electronics often referred to as "Mottronics". Achieving this resistive switching by minimal external field at room temperature is of paramount research and technological interest. Mott-IMT is often associated with structural modification, which is very important for optoelectronic and actuator applications. Here, we report a giant resistive switching ~900 % at room temperature in disordered polycrystalline $V_2O_3$/Si thin film stabilized at IMT phase boundary and associated structural transformation under small electric field. Increase of electron population in $a_{1g}$ band under field is responsible for Mott gap collapse that drives the structural transition. Furthermore, we also fabricated a room temperature Mott-FET with channel ON/OFF resistive ratio ~15. This study provides fundamental mechanism of the Mott-IMT in $V_2O_3$ as well as its device applications.**

Realization of room temperature electric field controlled volatile resistive switching in silicon compatible functional materials is highly desirable for low power device applications in optoelectronics, neuromorphic computation, actuators, sensors, logic devices etc. [1,2,3,4]. Insulator to metal transition (IMT) triggered by electric field in Mott insulators is believed to be a very promising way to control the resistive state. Additionally, Mott materials have a fundamental edge over the conventional semiconductors on the speed, charge carrier density, power consumption and dimensionality scaling [5,6,7]. $V_2O_3$ is a classic Mott material with strong electron correlation, which exhibits a temperature driven metal to insulator transition (MIT) around 160 K [8,9]. The electric or optical field induced IMT in $V_2O_3$ is very fast, (~nsec) and requires low energy making it a promising material for high-speed, low-power electrical and optical switching and modulation applications [1,10,11,12,13].

In Mott insulators, strong electron correlations may couple to lattice and spin degrees of freedom. Crystallographic *c/a* ratio plays a crucial role in stabilizing the ground state of such a material. Isostructural paramagnetic metal (PM) to insulator (PI) transition in $V_2O_3$ has been reported by tuning the *c/a* ratio through chemical doping, disorder or strain [14,15,16]. Reduction of the *c/a* ratio increases the orbital overlap leading to increase in the separation between the Hubbard bands that opens up a Mott gap resulting in an insulating state. This iso-symmetric MIT provides huge opportunities for new electronic devices, where the ultimate aim is to stabilize the system at the metal-insulator phase boundary, such that a minimal external trigger leads to a maximal resistivity change. Several correlated oxide/ ferroelectric hetero-structures have been studied previously using $V_2O_3$, $VO_2$, $Fe_2O_3$, $LaNiO_3$ and $NdNiO_3$ [17,18,19,20,21,22,23], however, their resistive switching is either less than 110% or at low temperatures. A resistive switching can also be achieved using electrolyte gating [24]. However, ionic-liquid modulation is not suitable for high-speed operation, device integration and in many cases induces an irreversible compositional change due to formation of oxygen vacancies at the liquid-solid interface [1,25]. Therefore, achieving high resistive switching by means of a small electric field at room temperature is of paramount research and technological interest.

To realize room temperature resistive switching, it is important to understand the fundamental mechanism of Mott IMT. In $V_2O_3$ it is still lacking. For example, Mott transition in many materials is accompanied by a structural phase transition (SPT). However, whether the crystal structure drives the Mott transition or electronic state drives SPT, is yet not clear. Here, with the help of simultaneous electric field dependent resistivity and Raman spectroscopy studies, we reveal that the electron-electron correlation is the key factor to drive the Mott transition in $V_2O_3$. Importantly, in this study the paramagnetic insulating phase (high resistive state -HRS) has been stabilized in polycrystalline $V_2O_3$/Si thin film at room temperature that can be transformed to the metallic state (low resistive state -LRS) by applying relatively small electric current (100mA) with a giant conductivity change ~ 900% without help of secondary drivers like ferroelectric interface. Further, for demonstrating its applications, we fabricated a prototype Mott Field Effect Transistor (MFET) with extremely low leakage current. This helped also to resolve the long standing fundamental concern whether the electric field induced IMT is of pure electronic origin or induced by joule heating.

## Results

**Thin film characterization**

The microstructure of $V_2O_3$/Si thin film grown by Pulse laser deposition (PLD) is investigated using scanning transmission electron microscopy (STEM), X-ray diffraction (XRD), and Raman spectroscopy (details in supplementary note S1). Fig. 1a & b show the cross-sectional high angle annular dark field (HAADF) STEM image and corresponding elemental mapping. High resolution STEM image (Fig. 1c) depicts the highly disordered nano-crystallites. Fig. 1d represent the fast Fourier transformed pattern of the area marked by the blue boxes in Fig. 1c, which show the different orientations of the nano-crystallites. The crystallographic *c/a* ratio obtained from the atomically resolved STEM-HAADF image (Fig. 1e) is smaller than the bulk and $V_2O_3$/$Al_2O_3$ thin film (table 1). Fig. 1f shows the schematic of *a-c* plane of $V_2O_3$ unit cell. XRD and Raman studies shown in Fig. 1g, h further confirm the phase purity and reduction of *c/a* ratio in $V_2O_3$/Si thin film (details in note S1).

Table 1: Structural parameters of $V_2O_3$ thin film obtained from XRD and TEM compared with bulk [26].

| Sample | *a* (Å) | *c* (Å) | *c/a* |
|---|---|---|---|
| Bulk | 4.953 | 14.011 | 2.829 |
| $V_2O_3$/ $Al_2O_3$ | 4.934+0.007 | 13.984+0.001 | 2.830+0.004 |
| $V_2O_3$/Si (XRD) | 4.974 ±0.002 (0.4% tensile) | 13.920±0.001 (0.6% compressive) | 2.798±0.001 |
| $V_2O_3$/Si (TEM) | 5.10 | 14.28 | 2.80 |

**Temperature dependence of resistance under different electric field**

Fig. 1i shows the resistance (R) vs temperature (T) behavior measured on the $V_2O_3$/Si thin film using linear four probe method under application of different electric fields (currents). A huge increase of resistance with decreasing T is observed under small probe current (10μA). The negative slope (dR/dT) indicates the insulating nature of the thin film even at room temperature in sharp contrast to the room temperature metallic phase in bulk and epitaxial $V_2O_3$/$Al_2O_3$ thin film (shown in Fig. S1) [27, 28]. Interestingly, with increase in the probe current, the change in resistance as a function of T becomes smaller and smaller (Details in note S2). The present results show that the room temperature insulating phase and resistive switching can be realized in undoped polycrystalline thin films with structural defects playing the role of Cr-doping [14].

Theoretical studies predicted reduction in Mott gap with a metastable state due to phase-coexistence and through electron injection collapses the gap under a very small applied electric field [29]. Defects and disorder in Mott materials often facilitate such phenomenon [11]. To check the possibility of a change in the electronic state driven by the electric field as well as the effect of electric field on associated crystal structure in the present disordered polycrystalline $V_2O_3$/Si thin film, we performed *in-situ* Raman and voltage (V)-current (I) characteristics at room temperature and low temperatures.

**Electric field induced resistance switching and crystal structure change at 300K and 7K**

A schematic of the *in-situ* measurements geometry is shown in Fig. 2a. Fig. 2b presents V-I characteristics at room temperature. Non-linearity in V-I characteristics indicate the electric filed induced resistive switching in $V_2O_3$/Si film. This is in sharp contrast to the metallic $V_2O_3$/$Al_2O_3$ film which shows linear V-I response at room temperature (shown in Fig. S1b). Corresponding differential conductance (DC) is plotted in the Fig. 2b. DC is obtained as DC = $\frac{\sigma_d(I) - \sigma_d(0)}{\sigma_d(0)} \times 100\%$., where $\sigma_d(I)$ is conductance, defined as $\sigma_d(I)$= $\left(\frac{dI}{dV}\right)_{I=I}$ and $\sigma_d(0)$= $\left(\frac{dI}{dV}\right)_{I=0}$. It shows a high DC change ~ 900% at room temperature on the application of small current (100mA). On the other hand, the metallic $V_2O_3$/$Al_2O_3$ thin film does not show any DC as shown in Fig. S1c. To probe the associated structural modification, we performed the Raman spectroscopy measurements. Fig. 2c shows the Raman spectra collected under different current values at 300K. Position of the $A_{1g}$ and $E_g$ Raman modes under zero field (254 and 236 cm$^{-1}$, respectively) are at higher wave numbers than observed in the metallic bulk and $V_2O_3$/$Al_2O_3$ thin film (242 and 210 cm$^{-1}$) [30] and is attributed to the lower *c/a* ratio in the $V_2O_3$/Si thin film [31]. For comparison, the Raman spectrum of metallic $V_2O_3$/$Al_2O_3$ thin film is also plotted in Fig 2c. The position of both the modes shows a shift towards the lower wave number and approaches to the metallic $V_2O_3$/$Al_2O_3$ film with increasing current. Thus, the Raman study indicates that on application of electric field the insulating R′ phase transforms to the metallic R phase. Therefore, these results demonstrate a unique way to tune the crystallographic unit cell by application of electric current which is the central factor for optoelectronic applications [1,2,32,33,34]. As discussed before, the $V_2O_3$/Si film is a room temperature paramagnetic insulator (HRS) having corundum structure (R′ phase) with lattice parameters slightly differing from the metallic state (LRS) corundum structure (R phase). With the application of electric field, the phonon modes show a softening without change in symmetry or undergoes isostructural HRS (R′) to LRS (R) phase transition originated from the strong electron-electron correlation (other possibilities are discarded in note S2).

Similarly, to probe the electric field effect on low temperature electronic and structural phase, we have performed in-situ Raman at 7K. With lowering T, the room temperature insulating R′ phase transforms to the insulating metastable structure with coexisting monoclinic $C_{2/c}$ and R′ phase and show a huge ~$2.2 \times 10^4$ times (highest ever to the best of our knowledge) DC switching under electric filed (see the note S3) and as a consequence the crystal structure is transformed to the R phase.

**Temperature dependent V-I characteristics and threshold field for IMT**

Fig. 3a presents the isothermal V-I characteristics at some selected temperatures. Detailed hysteresis behaviors are shown in Fig. S6 and discussed in note S5. A giant DC ($2.2 \times 10^4$ %) switching is observed at 7K and its value gradually reduces with increasing temperature (shown in Fig. 3b). Interestingly, the HRS to LRS occurs only after a threshold voltage value $V_{th}$ which decreases with increasing temperature (shown in Fig. 3c). The required threshold field ($V_{th}/d$) (15 V/cm at 300K) in our $V_2O_3$/Si thin film is at least two order smaller than the other prototypical materials like $AM_4Q_8$ (A- Ga, Ge; M - V, Nb, Ta; X - S, Se), $NiS_{2-x}Se_x$ and $V_{2-x}Cr_xO_3$ [35,36,37].

**Temperature dependent Raman measurement under zero and finite electric field**

To understand the thermal evolution of landscape of structural phase coexistence and effect of electric field on it, we performed T dependent Raman studies with and without application of external electric field. Fig. 4a represents the T-dependent Raman spectra without external electric field, which shows the onset of a structural phase transition from rhombohedral R′ (R-3c) to $C_{2/c}$ monoclinic phase at 150K ($T_{SPT}$) though the system remains in the insulator state and this phase transformation is not complete even up to 7K (shown in Fig. S5a). Both the phases coexist below 150 K, though monoclinic phase dominates over rhombohedral phase as the temperature is lowered. Interestingly, under the application of 100 mA current, the SPT is suppressed (shown in Fig. 4b) and the Raman spectra corresponding to the room temperature R phase persists even up to 7K. This is similar to our R vs T results under 100mA, where the room temperature electronic state is arrested and it remains in the LRS throughout the measured temperature window (Fig 1i).

**Room temperature Mott field effect transistor**

As shown in Fig. 1i a giant electric field induced conductance change at room temperature is observed in the present $V_2O_3$/Si film. This property can be utilized to fabricate a Mott FET (MFET). Apart from direct applications, such Mott transistor is used for confirmatory test about the mechanism involved in electric field induced IMT. If the IMT could indeed be triggered solely by electron injection on application of an electric field, then only resistive switching is possible in a three-terminal FET geometry with a high-k gate dielectric eliminating the leakage current. To check this, we fabricated a Mott FET transistor where the channel resistance in $V_2O_3$ is controlled by gate voltage. Schematic of the three terminal device structure is shown in Fig. 5a. The V-I characteristics of $V_2O_3$ channel are measured under different gate voltages (Vg) like (0→0.9V→0V→1.5V→0V) and plotted in Fig. 5b. For Vg = 0 V, the current is negligibly small, which can be considered as OFF state. However, on application of gate voltages (0.9 V, 1.5 V), the conductivity of the $V_2O_3$ channel increases significantly resulting in high current flow. This state can be considered as ON state. The channel conductance is plotted in Fig. S8. Because of very high dielectric gate insulation the leakage current is very small (~100 nA). On application of gate voltage, the insulator channel is converting to metallic state through Mott mechanism driven by electron injection making the channel a LRS. This result confirms that the HRS to LRS transition is driven by purely electronic Mott transition and rules out the possibilities of Joule heating. Notably this MFET requires very low power in comparison to the existing transistors.

**Discussions**

It is well known that the systems stabilized near the phase boundary can be switched to the other phase by small external stimuli and provides higher response. For example, high electromechanical response in relaxor ferroelectric is observed in systems stabilized by doping induced disorder near to the morphotropic phase boundary [38,39]. In present system, by controlling the disorder and defects, the $V_2O_3$/Si thin film is stabilized at metal–insulator phase boundary (for $c/a$ ~2.79) [14] which resulted in large resistive switching at room temperature.

In $V^{3+}$ valance state, two unpaired electrons belong to the 3d $t_{2g}$ orbital and because of the trigonal distortion, $t_{2g}$ orbital further splits into $a_{1g}$ and $e_g^\pi$ orbitals, and the spiting energy can be tuned by changing the crystallographic $c/a$ ratio [40]. Upon pressure [15], Cr doping [41,42,43] and epitaxial strain [14], the $c/a$ ratio can be decreased leading to a shorter V-V separation which causes a higher $a_{1g}$ ($d_{Z^2}$) orbitals overlap resulting in an increase in the coulomb repulsion energy. Because of this enhancement in coulomb repulsion, insulating state can be stabilized with reduced $c/a$ ratio.

Two-orbital Mott insulator model [29] predicted that within a metal-insulator coexisting region, the gap separating two bands of different orbital character can be collapsed by reducing the orbital polarization "m" (= $N_1$-$N_2$, i.e. population imbalance between the lower and upper orbitals). This orbital polarization can be tuned by application of electric field. G. Lantz et al [44] showed that the increase in the electron population in $a_{1g}$ electronic band (which reduces the orbital polarization) reduces the separation between $a_{1g}$ and $e_g^\pi$ bands. Therefore, when we apply electric field, the electron population in $a_{1g}$ band is increased through electron injection which decreases the $a_{1g}$- $e_g^\pi$ splitting and collapses the Mott gap above a threshold voltage $V_{th}$, which drives the HRS to LRS phase transition because of electron-electron correlation effect [29,45,46]. A schematic band diagram is shown in Fig. 5c,d. As a consequence of smaller separation between $a_{1g}$ and $e_g^\pi$ state, V-V separation increases which favors larger $c/a$ ratio [47] that leads to an isostructural R′ to R phase transition under application of external electric field (shown in Fig. 5e). Raman mode $A_{1g}$ represents the V atoms vibration along the $c$-axis and thus is highly sensitive to the V-V separation along $c$-axis [,44,47,48]. Therefore, $A_{1g}$ mode can be used to probe the trigonal distortion [49]. Softening of the $A_{1g}$ mode with external electric field signifies the increase in V-V distance and confirms the crystal structure transformation to the R phase. That is also responsible for the suppression of structural phase transition as well as huge resistance change with lowering of temperature. Once LRS phase stabilizes because of the electron injection in the $a_{1g}$ band under external field at room temperature, the gap collapsed state is maintained as long as the electron is populated in the $a_{1g}$ band. Therefore, the room temperature LRS- R phase is arrested under field even when the sample temperature is lowered to 7K. This is a possible mechanism of arresting the high temperature R phase under field cool cycle. Here, the electron injection is the driving force and clearly establishes that the electron-electron correlation is the main factor for the IMT in the $V_2O_3$.

The effect of electric field is to increase the electronic concentration in $a_{1g}$ band beyond the critical value ($V_{th}$) destabilizing the insulating state and triggering to a IMT. Since field induced carrier concentration increases with temperature, the switching voltage $V_{th}$ is reduced with temperature as shown in Fig. 3c. Notably the required $V_{th}$ for IMT is remarkably small, <1V. The effect of disorder/defects is like Cr doping, it reduces the $c/a$ ratio [31,49] and therefore, the potential barrier between two competing phases is reduced. As a matter of fact, in sample with higher defect densities, more carriers can be generated with less electric field or power [11,29,36,[50]]. For an optimal *c/a* ratio, the system can be stabilized at the HRS-LRS phase boundary and a very small external perturbation like electric field can trigger the IMT. Therefore, room temperature low power resistive switching can be achieved by controlling defect densities as observed in our $V_2O_3$/Si thin film and its multistate resistive switching in two terminal device configuration would be a potential candidate for the neuromorphic computing.

Realization of resistive switching in FET geometry at room temperature confirmed that the IMT is electron-electron correlation driven. The very small leakage current rules out the joule heating effect in the IMT. Here we showed 15 times ON/OFF resistivity ratio using $V_g$ =1.5 volt which would be a step towards the practical application of MFET as a low power device. However, more device optimization is needed to reach its full potential.

**Conclusion**

For applications, resistive switching mechanism compatible with Si is highly sought and in this study, resistive switching at room temperature is demonstrated in $V_2O_3$ polycrystalline films deposited on Si. It is shown that by controlling defects a phase boundary compound (film) with desired c/a ratio can be stabilized. Further, a three terminal test device mimicking Mott FET is fabricated and resistive switching is controlled through gate voltage. This device showed 15 times conductivity change in film at room temperature by application of small gate voltage which opens the possibility to fabricate correlated oxide based low power Mott transistor. The study unequivocally showed that the electron injection mechanism is responsible for the Mott insulator to metal transition along with isostructural transition in $V_2O_3$/Si film. Electric field induced conductivity along with crystal structure change in linear four probe geometry and FET geometry confirm that the Mott gap collapse is the key mechanism of the IMT.

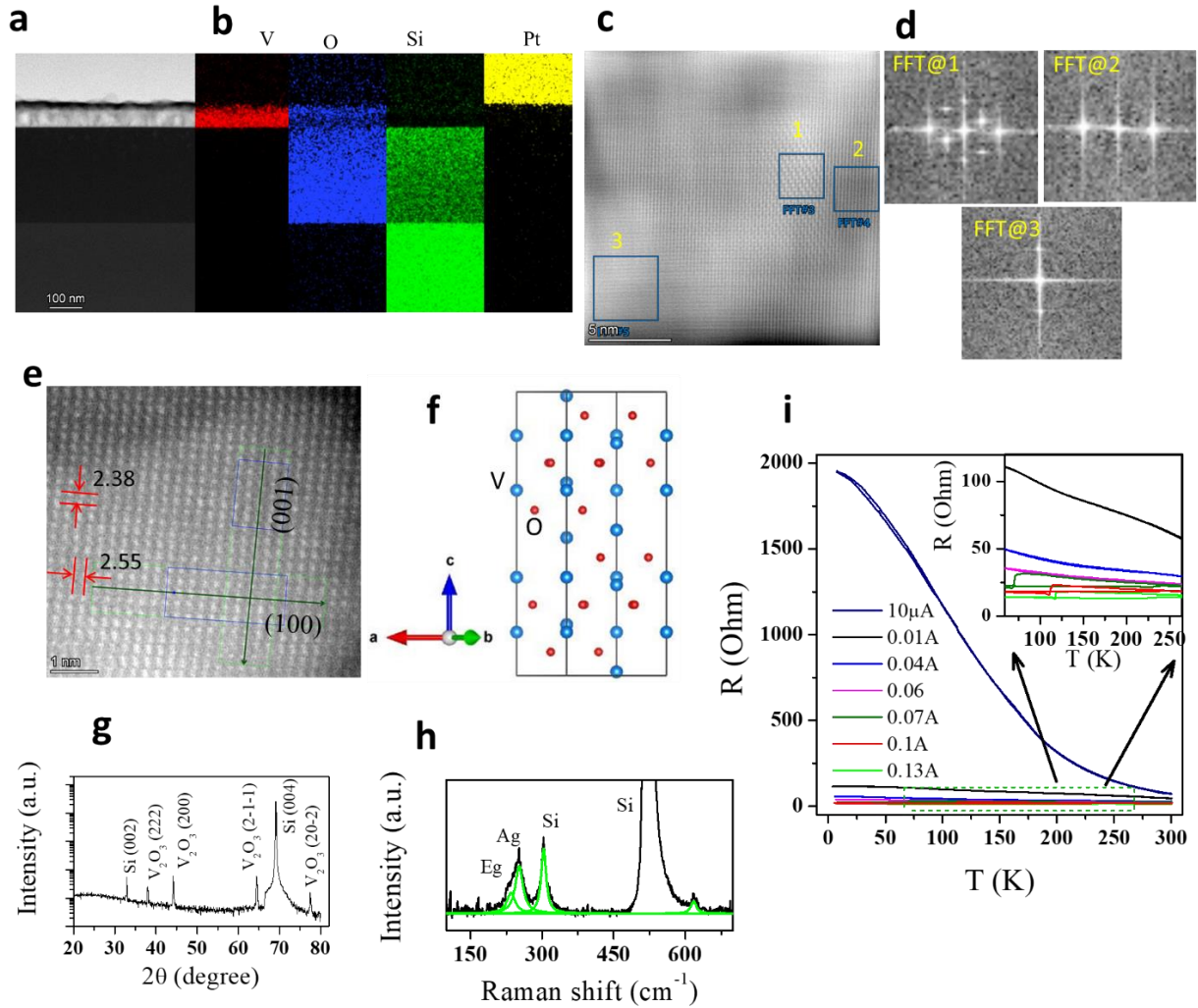

**Fig. 1| Characterization and electrical response of the V₂O₃/Si thin film. a** & **b,** Cross sectional STEM HAADF image and EDS mapping. **c,** Large area STEM image showing disordered regions. **d,** Fast Fourier Transformation of the selected area of Fig. **c**. **e,** Atomically resolved STEM image on *a-c* plane. **f,** V₂O₃ schematic. **g** & **h,** XRD and Raman spectra. **i,** Resistance R vs T under different dc currents from 10μA to 0.13A applied at 300K and resistance was measured in both heating and cooling cycles with temperature sweeping rate 3K/min. Inset shows the magnified view of the rectangular green dotted area.

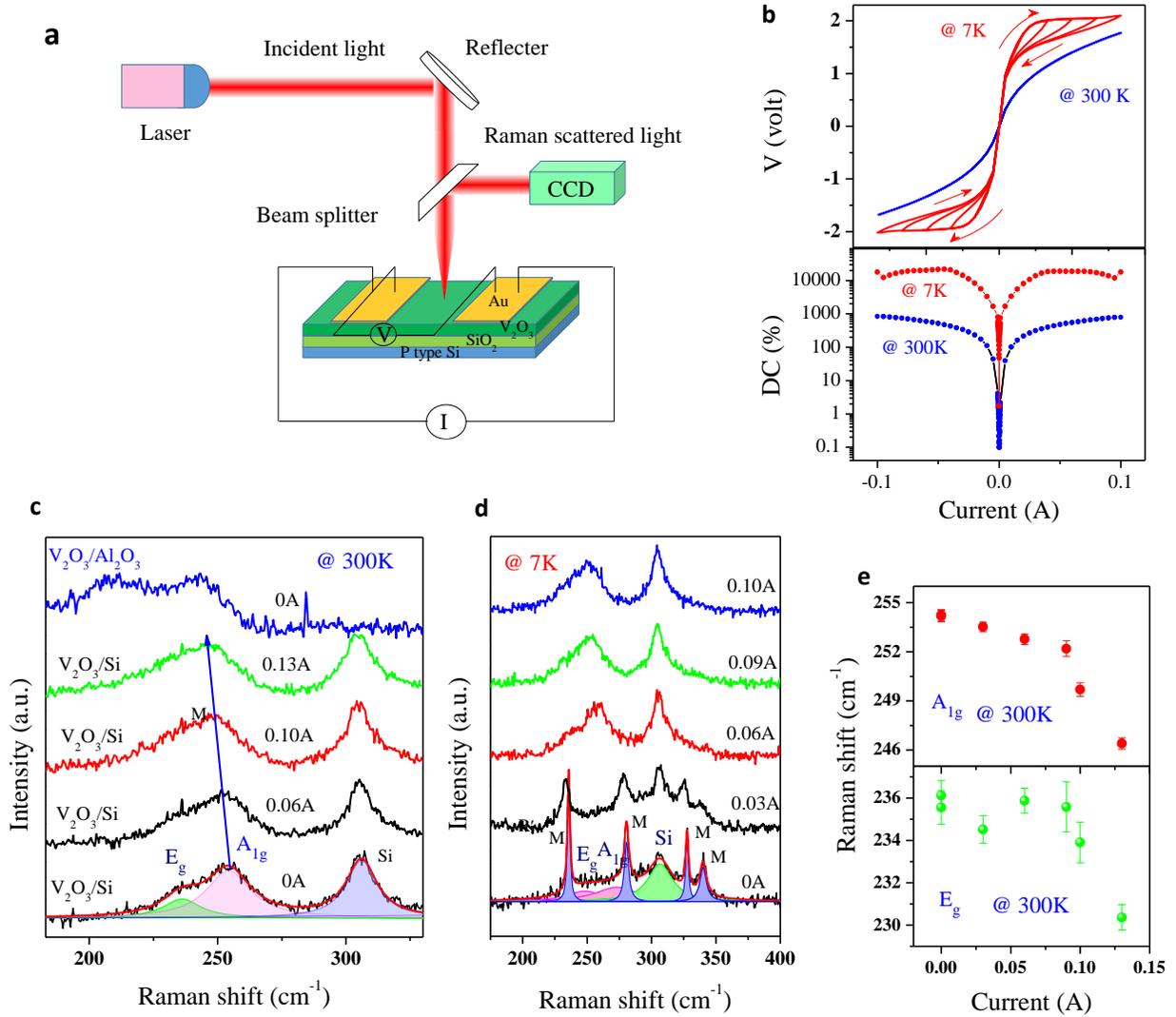

**Fig. 2| *In-situ* electric field dependent Raman and V-I characteristics of $V_2O_3$/Si thin film at 300K and 7K. a,** Schematic of the experimental geometry for simultaneous Raman and resistance measurements. **b,** V-I characteristic along with minor hysteresis loops and differential conductance (DC) as a function of applied electric current. **c,** Raman spectra at 300K, measured on $V_2O_3$/Si under different electric fields (currents) along with Raman spectrum of metallic $V_2O_3/Al_2O_3$ (top curve) shown for comparison. **d,** Raman spectra at 7K under field. Symbol M depicts the modes corresponding to the monoclinic phase ($C_{2/c}$). **e,** Variation of $A_{1g}$ and $E_g$ mode position (rhombohedral R′ phase) at 300K as a function of applied electric current.

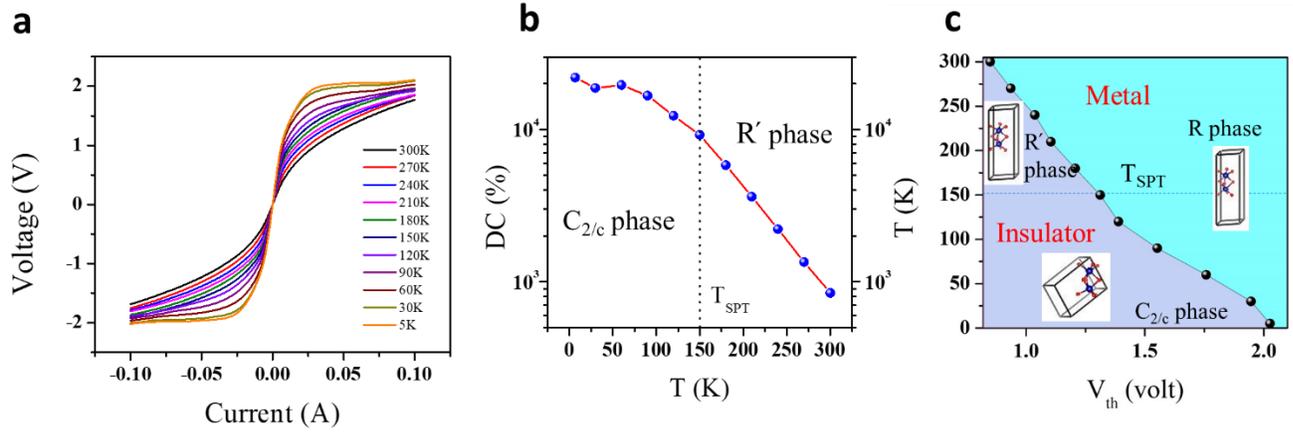

**Fig. 3| Electric field induced Differential Conductance (DC) change at different temperatures and IMT phase diagram of $V_2O_3$/Si thin film: a,** Temperature evolution of I-V characteristics of $V_2O_3$/Si thin film. **b,** Differential conductance change due to application of 100mA current at different temperatures. **c,** Phase diagram of the electric field induced IMT as a function of temperature. Threshold voltage $V_{th}$ of IMT decreases with increasing temperature. Separation between two electrodes "d" was 550μm. Blue dash line represent the onset temperature of structural phase transition ($T_{SPT}$). Crystal unit cells associated with the insulating and metallic phases are also shown. Both metallic R and insulator R′ phase belong to R-3c structure, but *c/a* ratio is higher in R phase. Monoclinic $C_{2/c}$ phase appears below $T_{SPT}$.

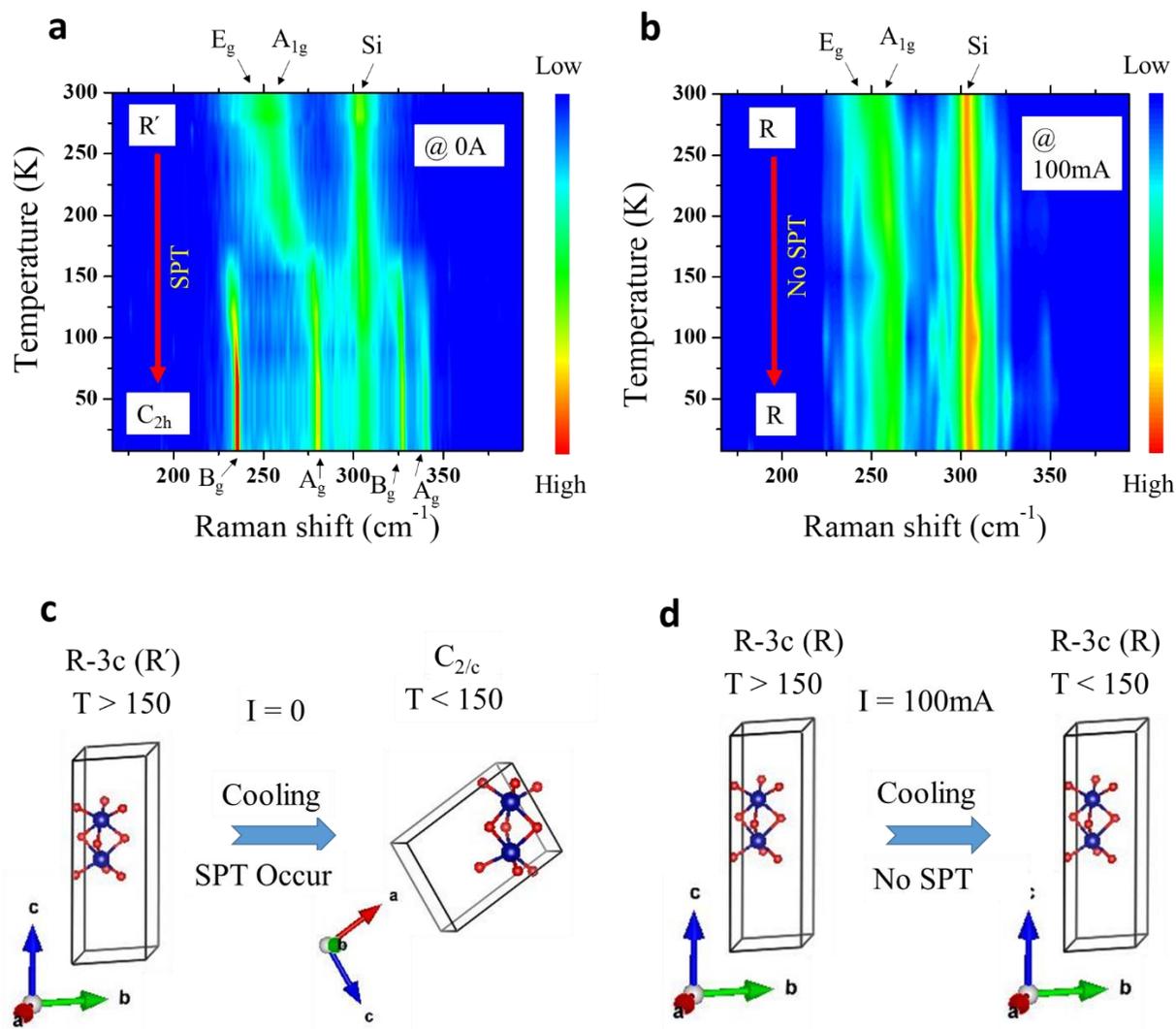

**Fig. 4| Temperature evolution of the microstructure under zero and finite electric field. a,** Contour plot of Raman spectra collected on $V_2O_3$/Si under zero current and **b,** under 100 mA dc current. **c & d,** Temperature evolution of the crystal structure measured under 0A and 100mA current cooled mode. Cartoon depicts R-3c crystal structural transformation to $C_{2/c}$ structure below 150K under zero current, and this transition is suppressed under 100mA current.

**Fig. 5| Mott field effect transistor characteristics and Mechanism of the electric field induced structural change in $V_2O_3$/Si thin film. a,** Schematic of Mott field effect transistor (MFET) structure. **b,** The channel I-V characteristics at room temperature under different gate voltages. **c,** Schematics of the energy band diagram in insulating and metallic state. **d,** Electronic band splitting under the trigonal distortion with different *c/a* ratio values. **e,** Two face sharing $VO_6$ octahedra corresponding to R′ and R phase.

## Materials and methods

**Growth and device fabrication:** The $V_2O_3$ thin film was deposited on 285nm $SiO_2$ coated (100) P type Si substrate using Pulse Laser deposition method (PLD). A sintered pellet of $V_2O_5$ was used as a target and 248nm KrF pulse Eximer laser of pulse energy $2J/cm^2$ and repetition rate 3Hz was used for deposition. Substrate was kept at a distance of 5cm from the target. During deposition substrate temperature was 650°C and the deposition chamber was evacuated to $10^{-6}$ mbar. After deposition the film was cooled down to room temperature with a rate of 2.2 K/min while maintaining the same pressure. 30 nm Au electrodes were deposited on $V_2O_3$ thin film using magneton sputtering method for four probe resistivity measurements. For bottom gated three terminal device, we stacked the P type Si with Cu using Silver paste. The device structure is presented in Fig. 5a.

**Structural characterization:** The XRD $\theta$-$2\theta$ and X ray reflectivity (XRR) measurements were performed using PAN analytical X'PERT high resolution x-ray diffraction (HRXRD) system equipped with a Cu anode. Raman measurements were performed using Horiba Jobin-Yvon, France, micro Raman spectrometer equipped with a 633 nm excitation laser, 1800 g/mm grating, an edge filter for Rayleigh line rejection and a CCD detector giving a spectral resolution of ~1cm$^{-1}$. The Laser was focused to a spot size of ~1 micron onto the surface of the sample using a 50× objective lens. The spectra were collected in back scattering geometry. A liquid helium flow cryostat from Janis (USA) was used for cooling the sample temperature. In situ resistivity and Raman measurements at different T were performed using this cryostat.

**TEM:** STEM images were collected using aberration (probe) corrected Titan Themis 300 operated at 300kV equipped with SuperXG quad EDS detector. 24 mrad convergence angle used for EDS and HAADF collection angle of 48-200 mrad were used at a camera length of 160mm.

**Resistivity measurements:** Resistivity and I-V characteristics were performed in the linear four point probe configuration using Keithley 2425 source meter and Keithley 2002 multimeter. Temperature was controlled using the Lakeshore 335 temperature controller. Temperature dependent resistivity measurements in cooling and heating cycles were performed in the temperature sweeping mode with 3K/min temperature sweeping rate. The resistance measurements under different current were performed by applying some specific constant dc current and the corresponding voltage drop was measured. Isothermal I-V characteristics were obtained in continuous increment/decrement of current and measuring the corresponding voltage difference between two probing points. For measuring the isotherm at a particular temperature, each time the sample was cooled from 300K with a cooling rate 5K/minute. After reaching the desired temperature, isotherm was obtained using the sequence 0A→0.1A→0A→-0.1A→0A. The sequence of minor hysteresis measurement is shown graphically in Fig. S2. Electrodes separation was 550μm.

**Acknowledgement**

Binoy De and S. B. Roy acknowledge financial support obtained from the Raja Ramanna Fellowship program sponsored by DAE, Government of India. Authors acknowledge the Advanced Facility for Microscopy and Microanalysis (AFMM), Indian Institute of Science, Bengaluru, India for STEM measurements.

**Author contributions**

B. K. D, V.G.S and S. B. R designed the experiments. B.K.D and H. S. K prepared the sample. S.K.P and N. K performed STEM measurements and analyzed the data. B.K.D, Divya and P.S designed the in-situ Raman and resistivity setup, and performed the measurements. B.K.D, S. B. R and V.G.S wrote the paper with input from all authors. All authors discussed the results and contents of the paper.

**Competing interests**

The authors declare no competing interests.